# Ultrasonic Delamination Based Adhesion Testing for High-Throughput Assembly of van der Waals Heterostructures

Tara Peña,*,† Jewel Holt,† Arfan Sewaket,† and Stephen M. Wu†,¶

†*Department of Electrical & Computer Engineering, University of Rochester, Rochester, NY, USA.*

¶*Department of Physics & Astronomy, University of Rochester, Rochester, NY, USA.*

E-mail: tpena@ur.rochester.edu, stephen.wu@rochester.edu

## Abstract

Two-dimensional (2D) materials assembled into van der Waals (vdW) heterostructures contain unlimited combinations of mechanical, optical, and electrical properties that can be harnessed for potential device applications. Critically, these structures require control over interfacial adhesion in order for them to be constructed and to have enough integrity to survive industrial fabrication processes upon their integration. Here, we promptly determine the adhesion quality of various exfoliated 2D materials on conventional $SiO_2$/Si substrates using ultrasonic delamination threshold testing. This test allows us to quickly infer relative substrate adhesion based on the percent area of 2D flakes that survive a fixed time in an ultrasonic bath, allowing for control over process parameters that yield high or poor adhesion. We leverage this control of adhesion to optimize the vdW heterostructure assembly process, where we show that samples with high or low substrate adhesion relative to each other can be used selectively to construct high-throughput vdW stacks. Instead of tuning the adhesion of polymer stamps to 2D materials with constant 2D-substrate adhesion, we tune the 2D-substrate adhesion with constant stamp adhesion to 2D materials. The polymer stamps may be reused without any polymer melting steps, thus avoiding high temperatures (<120°C) and allowing for high-throughput production. We show that this procedure can be used to create high-quality 2D twisted bilayer graphene on $SiO_2$/Si, characterized with atomic force microscopy and Raman spectroscopic mapping, as well as low-angle twisted bilayer $WSe_2$ on h-BN/$SiO_2$/Si, where we show direct real-space visualization of moiré reconstruction with tilt angle-dependent scanning electron microscopy.

# Introduction

Two-dimensional (2D) materials - such as graphene, hexagonal boron nitride (h-BN), and transition metal dichalcogenides (TMDs) - exhibit a vast range of intriguing (opto)electronic properties. Because of the exceptionally weak out-of-plane van der Waals (vdW) interactions, these 2D materials can be mechanically isolated then layered into vdW heterostructures. This vdW heterostructure assembly procedure typically involves dry-transfer processes that rely on temperature tunable adhesion between 2D materials and polymer stamps. The success of the vdW heterostructure assembly process depends critically on the control over the relative adhesion of 2D materials to stamps, substrates, and other 2D materials, leaving adhesion as the most important process parameter to control in 2D vdW stacking. High-throughput, simple, and quick tests of relative adhesion are therefore highly important for the optimization of any vdW heterostructure assembly process and may open doors to new techniques that may supplant the current status quo methods.

Interfacial adhesion has been experimentally investigated through a number of experiments such as interactive scanning probe tips[1], pressurized blister tests[2], and buckle delamination tests[3]. All of these works paved the way for quantitatively obtaining interfacial adhesion energies between 2D materials and various 3D-bonded substrates. However, because this parameter entirely depends on the interaction between two interfaces, it varies easily with the cleanliness of the substrate or environmental conditions[4]. Additionally, since these are specialized tests, they do not allow for a simple high-throughput mechanism to determine if certain process parameters are contributing to these variations in adhesive properties.

In this work, we present ultrasonic delamination threshold testing as a method to infer the interfacial adhesion quality between 2D materials and their respective substrates. Ultrasonic cleaning is a standard cleaning technique, where ultrasonic sounds waves are generated to agitate fluids and subsequently any submerged samples. This aggressive cleaning technique has been used previously to corroborate quantified adhesion measurements of polymer/substrate interfaces[5]. While this method does not yield a direct quantitative measure of 2D material/substrate interfacial adhesion energies, it can be used to judge relative adhesion between two different samples in a simple, quick, and reliable manner. Using this method of adhesion testing, we can judge adhesion quality between exfoliated graphite flakes and a conventional $SiO_2$/Si substrate under different process conditions, and then use this knowledge to intentionally fabricate graphene flakes with high and poor adhesion to the $SiO_2$ surface. This allows for direct control over an adhesion variable that otherwise would not be accounted for in the vdW heterostructure assembly process.

Using this control over substrate adhesion, we show that we can optimize the throughput of vdW heterostructures and side-step many of the undesirable constraints set by the standardized procedure. Here, we choose to vary adhesion by exfoliating 2D flakes on plasma-treated $SiO_2$/Si substrates that vary with ambient exposure. By varying between high and low adhesion at the 2D material/substrate interface, we can control which 2D flakes can be delaminated by the polymer stamp during vdW heterostructure construction. Moreover, we find when a 2D flake on the polymer contacts a highly-adhered flake on $SiO_2$/Si, the 2D flake on the polymer will prefer to remain on the highly-adhered flake because of the high adhesion to the substrate and the additional vdW interactions between the two layers, therefore constructing a vdW heterostructure directly on $SiO_2$/Si. This process represents the reverse process of the standard dry-transfer technique, where the flakes remain on the polymer stamp after contact and transfer to a $SiO_2$/Si substrate requires a polymer melt. This new method allows for vdW heterostructures to be constructed directly onto the substrate and the polymer stamps may be reused. Melting the polymer is disadvantageous since this may need high-temperatures (~200°C) and will inevitably lead to residues. To combat these polymer residues, the entire sample must go through rigorous solvent baths and/or a h-BN encapsulation is required to protect the interface of interest underneath. Directly constructing the vdW heterostructure onto a target substrate significantly enhances the throughput of this process, furthermore the structure quality is retained without the need of extensive post-processing steps to clean polymer residues.

Using this new high-throughput vdW heterostructure assembly process, we can quickly produce several TBG samples varying in twist angle, then probe sample quality with atomic force microscopy (AFM) and Raman spectroscopy. Finally, we show that this procedure can be extended to other 2D materials by constructing twisted bilayer $WSe_2$ on h-BN/$SiO_2$/Si (with high adhesion engineered between h-BN and $SiO_2$ interface) and show direct real-space imaging of moiré reconstruction through tilt-angle dependent scanning electron microscopy. These results show us that our adhesion testing techniques and new vdW heterostructure assembly process can reliably generate high-quality samples, by reproducing various important effects currently at the forefront of exploration in 2D heterostructures constructed with vdW assembly[6].

## Methods

All $SiO_2$/Si substrates are placed in an acetone then isopropanol ultrasonic baths for 15 minutes each, then dried with $N_2$. Subsequently, the substrates are oxygen plasma cleaned at 100 W and 250 mtorr for 3 minutes. The reactive ion etching chamber is also oxygen plasma cleaned for 30 minutes prior to putting any substrates inside, this ensures no contaminates can be redeposited onto the $SiO_2$ surface. The graphene flakes are produced with the scotch tape method onto pre-treated 300 nm $SiO_2$/Si substrates, then the substrates are heated at 100°C for 90 s with the graphite tape still in contact. The tape is slowly removed after the heating procedure is completed and the substrate has cooled. For monolayer $WSe_2$, we use Gel-Pak polydimethylsiloxane (PDMS) exfoliation to produce large area monolayers, then these monolayers are transferred onto 90 nm $SiO_2$/Si substrates. Both graphene and $WSe_2$ monolayers are identified with optical microscopy, which we thoroughly confirm prior to vdW heterostructure assembly with AFM and Raman spectroscopy[11].

## Results & Discussion

For the delamination testing, we fabricate numerous exfoliated graphite flakes on plasma-treated $SiO_2$/Si substrates that vary in adhesion. Oxygen plasma treatment is commonly used to clean $SiO_2$/Si substrates before exfoliation to increase monolayer flakes from a few $\mu m^2$ to well over 100 $\mu m^2$ in area[7]. Oxygen plasma treatment on the $SiO_2$ surface prior to exfoliation removes hydrocarbon contaminants and environmental adsorbates that pre-exist on the surface, additionally the plasma increases the surface reactivity of the $SiO_2$ (Fig. 1a,b)[8]. This means that the highest adhesion can be achieved by directly exfoliating the 2D materials onto the plasma-treated surface immediately after taking the substrate out of the chamber, ensuring the cleanest interface between the 2D material and substrate and high surface reactivity of the substrate's surface (Fig. 1c). The longer the $SiO_2$ is re-exposed to ambient conditions, the more environmental adsorbates are redeposited onto the surface and limits interactions with the exfoliated 2D flakes (Fig. 1d)[4,9]. The adhesion energies between monolayer graphene and $SiO_2$ varies tremendously in the literature between 100 and 450 mJ/m$^2$ in the literature[2,10], which is possibly a manifestation of these discrepancies with substrate cleaning and ambient exposure. We choose to control adhesion by varying the time the plasma-treated substrates are in ambient prior to exfoliation.

To observe this variation in graphite adhesion to plasma-treated substrates, we conduct ultrasonic delamination threshold testing. Here, we take the several graphite/$SiO_2$ samples that vary in the amount of time the plasma-treated $SiO_2$ was left in ambient prior to exfoliation between 1 minute and 60 minutes. We image several regions with graphite flakes over the samples (Fig. 1f,h), then place them into an IPA ultrasonic bath for 5 minutes, and finally repeat imaging over those exact regions (Fig. 1g,i). From this procedure, we find almost 90% of the graphite flakes survive the ultrasonic bath when the exfoliation is immediate (< 1 minute), which falls off quickly to only 15% after a 60 minute ambient exposure time (Fig. 1e). These results suggest that adhesion is changing rapidly with the substrates' exposure to ambient, more importantly we can infer that exfoliating under 1 minute and after 60 minutes of ambient exposure will guarantee either graphite flakes with high or low adhesion respectively. We can clearly observe how fixed graphite flakes are preserved from the ultrasonic bath test (Fig. 2a,b), while the free graphite flakes almost entirely are no longer on the substrate after (Fig. 2c,d).

Using a free and readily available image analysis software (ImageJ), we can quantitatively extract the area of graphite flakes that survive the ultrasonic bath testing presented in Fig. 1e. The optical micrographs of these areas were converted into binary images using ImageJ, where the substrate is highlighted in white and the graphite flakes in black (Fig. 2). To construct the binary images, we utilize two methods (Fig. 2a,b and Fig. 2c,d). Fig. 2a,b presents a method which outlines the edges of the graphite flakes, through an edge detection algorithm (a plugin provided by ImageJ). The algorithm calculates intensity gradients throughout the original image (Fig. 2e), where the gradient values are presented in Fig. 2a. Once the graphite edges are identified from the calculated gradient values, they are "filled in" to construct the final binary image in Fig. 2b. Fig. 2c,d presents a second method, where image thresholding is employed instead. Most graphite flakes within this thickness range have RGB values below the substrate value[11], therefore the background (substrate) of the original image can be brought to saturation (RGB = 255) by increasing the overall RGB value of the image (Fig. 2c). If any of the graphite flakes have RGB values above that of the substrate mean, the overall image's RGB value can be decreased such that the substrate is all black (RGB = 0), then the image can be inverted. Once the background is saturated, image thresholding will similarly yield a binary image where the graphite flakes are highlighted in black (Fig. 2d). Finally, these two images can be overlaid to extract the final binary image (Fig. 2f), two methods are used to affirm most of the exfoliated graphite features in the optical micrograph are accounted for. We also note, because monolayer graphene's contrast is extremely close to that of the substrate (|$RGB_{1L}$ - $RGB_{substrate}$| < 5), these methods are optimized to specifically capture these monolayer graphene features. Using this

procedure, quantifying the area of 2D materials across a sample can be quickly completed with high-accuracy.

To validate the results we observed from the ultrasonic delamination testing are from adhesion alone, we construct high adhesion (fixed) and low adhesion (free) graphene flakes on separate $SiO_2$/Si substrates that vary with their respective exposure times. Next, we bring PC film on PDMS dome stamp in contact with these respective graphene samples to attempt to delaminate them from their respective substrates. When the PC makes contact on the $SiO_2$ substrate, the substrate temperature is ramped up to 120°C, then the PC will expand over the targeted graphene flake and is kept in contact for five minutes. After the five minutes, the temperature is ramped down by 5-10°C every 60 seconds until either the substrate temperature is approximately at 70°C or until the PC retracts. When the PC contact pulls back over the graphene flakes, we find that the PC can only pick up the graphene samples that have engineered poor adhesion (free graphene flakes) while the highly adhered flakes entirely remain on the substrate. Thus, under constant 2D-stamp adhesion conditions, flakes with relatively lower (higher) substrate adhesion may remain free (fixed) to the substrate, yielding an opportunity to leverage relative adhesion as an advantage in vdW heterostructure assembling.

After creating these graphene samples with engineered substrate adhesion, we use them to construct high-throughput TBG on $SiO_2$/Si. After constructing free graphene samples, we can pick-up this layer with the standard PC/PDMS process as described previously (Fig. 3a,d). Once the free graphene is picked up by the PC/PDMS stamp, a fixed graphene sample can then be placed and fixated underneath the stamp (Fig. 3b,e). The two graphene layers are optically aligned to each other as desired, then the stamp is brought into contact with fixed graphene/$SiO_2$ surface. We find when the free graphene is laid over the fixed graphene, the free graphene will delaminate from the stamp onto the fixed graphene's surface permanently and leave a TBG/$SiO_2$/Si behind (Fig. 3c,f). This suggests that the fixed graphene/$SiO_2$ adhesion is substantially greater than the graphene/PC interfacial adhesion, and the additional vdW interaction between the two graphene layers triggers the ability for the free graphene to stay on the fixed graphene. This is the reverse process of typical vdW heterostructure constructions, where the 2D material adhesion to the substrate is always low and flakes are picked up by the more adhesive stamp. We reemphasize that this mechanism to create TBG/$SiO_2$/Si is quite favorable since the PC does not need to be melted like in other processes, meaning that the PC/PDMS stamp can also be reused for several iterations and this structure does not suffer from polymer residues. In this way, one could have wafers of "free" and "fixed" 2D monolayers and quickly assemble them into the desired vdW heterostructure. This procedure is also desirable for temperature sensitive materials, and/or vdW heterostructures that are intended to be examined with scanning probe experiments.

Another improvement provided by this process is this can easily yield bubble-free TBG structures by simply adjusting substrate temperature. In this particular structure, interfacial air blisters can only be formed in between the two graphene layers, therefore contacting the free graphene over the fixed graphene is the most crucial step in order to achieve a flat interface. We have found heating the sample to 100°C minimally during this step will reliably allow no air to be trapped in the interface. This temperature allows for air to be mobile, while the dome stamp provides enough force and direction to drive the bubbles out of the interface. To confirm this is the temperature needed, we attempted transfers below 100°C (Fig. 4a). We provide an example of a TBG structure where contact was conducted at 50°C, here interfacial air clearly detectable via differential interference contrast (DIC) micrographs (Fig. 4a) and AFM (Fig. 4b). Conversely, when this step is conducted at 100°C, a smooth interface be optically observed (Fig. 4c) and confirmed again in AFM (Fig. 3d). We do not observe any introduced residues from the process since the average surface roughness on the TBG region is precisely the same as that of the plasma cleaned $SiO_2$ surface alone ($R_a$ ~ 0.09 nm). We also note that we generally maintain PC contact speeds below 1 μm/s, since this reduces the possibility of the free graphene layer from folding or wrinkling.

In order to engineer twist angle with this method, we resort to aligning the zigzag edges of the graphene monolayers. After aligning the zigzag edges of the free and fixed monolayers, one can then rotate (by the desired twist angle) the fixed graphene/$SiO_2$ substrate before placing the free graphene monolayer. Here, the high-throughput allows us to produce many samples quickly and increase the probability we achieve the desired twist angle. We then perform Raman spectroscopic mapping on these TBG/$SiO_2$/Si structures, to inspect if the intended twist angle was achieved. Unique to high-angle TBG structures, an additional phonon mode (termed as the R' and R bands) arises from new scattering paths that are created by the moiré superlattice[12,13]. This moiré-activated phonon mode can be extracted from phonon dispersion curves along the $\Gamma$-$K$ direction in bilayer graphene, where the longitudinal and transverse optical phonon branches are assigned to the R' (θ < 10°) and R (θ ≥ 10°) peak positions respectively[14]. Thus, we then perform Raman spectroscopic mapping to verify the twist angle in TBG structures (WiTec Alpha300R Confocal Raman microscope, 532 nm excitation laser, ~4 mW power, ~0.7 μm spot size, and 1800 l/mm spectrometer grating). We first fabricate TBG structures ranging in twist angles from 6-22° using the edge alignment process discussed above. In Fig. 5a, we display Raman spectra from TBG samples with the intended engineered angle, where a successful sample is determined from having the expected R' (R) - band peak position within ±1° degree accuracy (denoted in Fig. 5a with asterisks).

Between 5° and 9°, the R'-band position is ~1625 $cm^{-1}$. In this regime, the intensity ratio between the G-peak and R'-peak ($I_G$/$I_R$) and R' peak position vary with twist angle (Fig. 5a). At 9°, the superlattice bands appear both roughly at 1620 and 1515 $cm^{-1}$, then larger angles have the only one R-band that linearly redshifts (approximately 1470 and 1395 $cm^{-1}$ for 14° and 22° respectively). Additionally, none of the samples below 20° display the defect peak (denoted as the D-band) in graphene at ~1340 $cm^{-1}$, neither in the TBG region or the individual monolayer regions[15], confirming the quality of the exfoliated monolayers is preserved. Above 20°, a D-like band at 1350 $cm^{-1}$ is expected to appear[13] that is unrelated to flake damage, which is observed in the inset of Fig. 5a. This D-like band appearing from the superlattice is supplemented by the fact this band does not appear in either of the monolayer regions of the sample. We are also able to observe other superlattice effects such as the G-band enhancement for TBG samples varying between 10° and 15°, which is fully consistent with other works and attests to the high quality of the new process.

Finally, we analyze Raman spectroscopic mapping to confirm the homogeneity of the G-band and R'-peak positions (Fig. 5c,d). Here we can observe that the TBG structure is mostly uniform. However, there are detectable variations in strain and doping across the entire sample. Strain is a ubiquitous consequence from vdW stacking and locally alters the twist angle in in fabricated 2D moiré superlattices.

Particularly, heterostrain (inequivalent strain between layers) can substantially modify any superlattice periodicity with small amounts of strain magnitude[16]. To corroborate this fact, we can plot the G-peak and 2D-peak (secondary peak at ~2685 cm$^{-1}$) frequencies across the sample to investigate variations with strain and carrier concentrations (Fig. 5b)[17]. In Fig. 5b, we observe a strong linear correlation between these two peaks following the strain axis (2.2 dashed line), indicating the presence of strains across the sample. This level of strain variation in our TBG sample is quite on par with other high-quality graphene heterostructures[18–20], and this effect is unlikely due to unique variations from our proposed process.

To demonstrate the adaptability of this procedure, we now extend this technique to construct a different 2D moiré superlattice: twisted bilayer $WSe_2$ on h-BN/$SiO_2$/Si. Because we have already investigated high-angle structures, we next attempt to use this procedure to create low-angle structures that require much more precision. To achieve this level of twist angle precision, we decide to tear and stack a large monolayer $WSe_2$ with the PC/PDMS stamp alone (analogous to tear and stacking with a h-BN mask)[21]. Using this reverse stacking process, we can quickly create twisted bilayer $WSe_2$ structures on the polymer and reliably drop them off onto highly adhered h-BN flakes. To achieve this, we similarly make sure that the exfoliated $WSe_2$ flakes have poor adhesion to the $SiO_2$ surface by leaving the substrates in ambient for over an hour. Once a large enough monolayer $WSe_2$ flake (> 20 µm$^2$) is on the $SiO_2$ and identified, the stamp is slowly brought in contact with the substrate, then the PC is ensured to only cover half of the monolayer during the entire heating and cooling pickup process. We note that the dome-shaped stamp is critical to reliably tear and stack the monolayer (without the typical h-BN mask) since this shape allows better control of the PC contact area. We find in this way; the monolayer can be broken in half and be picked up by the stamp. Therefore, we can then stack the monolayer back onto itself, and pick up a twisted bilayer $WSe_2$ on the stamp. Initially, we do not induce a twist angle between the two $WSe_2$ halves, this decision is common to construct exceptionally long moiré periods (hundreds of nm) since minor variations will always exist that trend away from the exact 0° (3R) twisted structure (monodomain).

Now that a twisted bilayer $WSe_2$ is fabricated on the stamp, we then place a h-BN/$SiO_2$/Si structure underneath where high adhesion is engineered between the h-BN/$SiO_2$ surface. Therefore, when the twisted bilayer $WSe_2$ on the stamp contacts the highly adhered h-BN, it again prefers to stay on h-BN and leaves a twisted bilayer $WSe_2$/h-BN/$SiO_2$/Si structure with a twist angle ~0° (Fig. 6a). From this procedure, we can fabricate substrates with free $WSe_2$ monolayers and fixed h-BN flakes to create such structures in a high-throughput fashion. For twist angles below 1°, atomic reconstruction becomes quite pronounced which enables discrete triangular AB and BA domains to form (Fig. 6b)[22]. We choose to resolve this reconstructed moiré superlattice and potential air bubbles simultaneously with scanning electron microscopy (SEM). Tilt-angle dependent SEM has recently been seen to allow a lattice-mediated channeling dependence of secondary electron emission. Once the sample is imaged at an angle with respect to the incident beam, each lattice configuration will independently have optimized channeling parameters depending if the incident beam is parallel (and "channels") to that specific lattice configuration (Fig. 6c), allowing contrast differences between stacking orders to be resolved across the same sample with this technique[23]. Since reconstructed superlattices have discrete AB and BA domains, these domains can be resolved with angle-dependent SEM imaging and thus be able to image the moiré periodicity directly.

Employing this technique, we can directly image the marginally twisted bilayer $WSe_2$/h-BN/$SiO_2$/Si structures we fabricate with this modified dry-transfer process (Zeiss Auriga Scanning Electron Microscope, accelerating voltage of 0.5 kV, a 30 μm aperture, tilt angle of 40°, and with a standard SE2 detector). On average with this method of tear and stacking, we can reliably obtain ~500 nm moiré periodicities ($\theta \sim 0.038°$). With this procedure, we can obtain several samples with a few micron-squared areas without bubbles and exhibit locally uniform reconstructed moiré domains (see Supplementary Information), which is comparable to the quality produced by other works constructing marginally twisted bilayers[23,24]. We note that this procedure does not perfectly yield bubble-free twisted $WSe_2$ throughout the entire sample, like presented in the TBG structures. We believe the issue with air being trapped occurs when the polymer contacts over the thick h-BN, it will snap over this height difference, therefore not allowing a steady speed of < 1 μm/s. Most likely, this can be also remedied by using contacting over thinner h-BN flakes. However, the quality of these vdW heterostructures are still similar with those presented in the literature and does not present a limitation unique to our proposed process. The ability to procedure high-quality structures with at a much faster pace, coupled with precisely tear-and-stacking these monolayers, is an exceptional combination of advantages to have to continuously create any set of twisted vdW heterostructures compatible with any experimental setup.

## Conclusion

We have presented a high-throughput method to infer the quality of adhesion between 2D materials and their respective substrate, allowing us to present a much more complete vdW heterostructure assembly procedure that is optimized for speed while retaining quality. Judging then engineering adhesion is the crux to substantially optimizing the vdW heterostructure assembly process, where we can use this enhanced control over substrate adhesion to avoid polymer residues, allow h-BN encapsulations to be optional, and reduce the overall required temperature during the process. Our process capitalizes on engineered 2D material adhesion to their respective substrate, instead of only tuning the polymer stamp adhesion, which allows the vdW heterostructure to be directly fabricated onto the target substrate. This process enhances the throughput and reliability of such high-quality moiré structures on $SiO_2$/Si substrates, for both research and mass-production purposes. Because these twisted vdW heterostructures exhibit novel moiré-dependent properties, such as correlated electron phenomena[25–27], interlayer moiré excitons[28,29], ferroelectricity[24,30], new magnetic ground states[31], etc, investigating the intrinsic properties of any vdW heterostructure crucially relies on high-throughput fabrication processes that reliably yield high-quality samples. This proposed vdW procedure is compatible with any 2D material system and the finalized structure can be tailored for any experimental setup, all while enhancing the production speed. Learning about the processes which contribute to high or low substrate adhesion allows for continued optimization of the vdW heterostructure assembly process, where this new high-throughput vdW assembly method may be combined with robotic assembly tools to further enhance reliability and throughput of the process in an generalized manner for all 2D materials[32].

## Conflict of Interest

The authors have no conflicts to disclose.

## Data Availability

The data that support the findings of this study are available from the corresponding author upon reasonable request.

## Acknowledgment

We wish to acknowledge support from the National Science Foundation (OMA-1936250 and ECCS-1942815) and the National Science Foundation Graduate Research Fellowship Program (DGE-1939268). This work also made use of the Cornell Center for Materials Research Shared Facilities, which are supported through the NSF MRSEC program (DMR-1719875). We also acknowledge Andrey Sushko for helpful discussions on SEM imaging.

## Supplementary Material

More details on ultrasonic delamination testing, PC/PDMS dome stamp fabrication, and transfer process.

# Figures

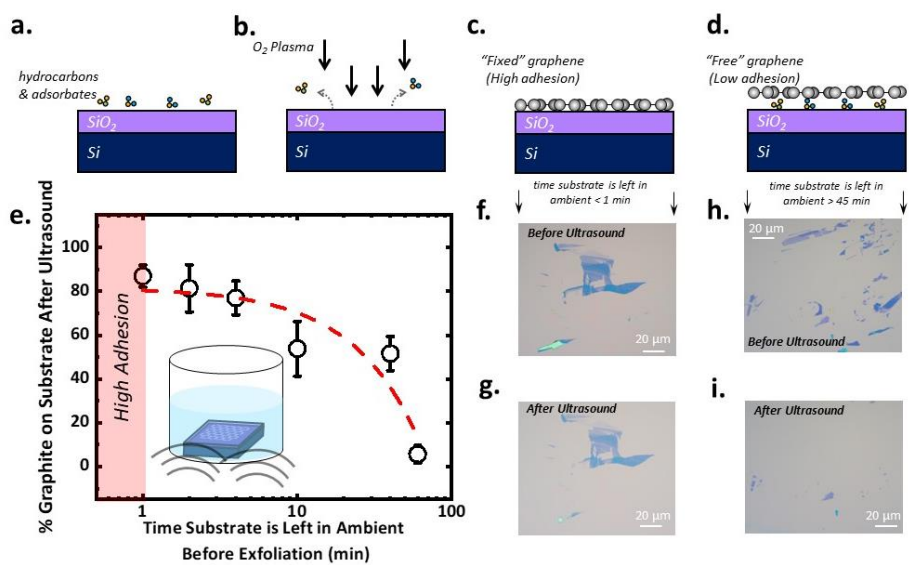

**Figure 1: (a)** A SiO$_2$/Si substrate in ambient conditions, with a layer of hydrocarbons and adsorbates. **(b)** Schematic of oxygen plasma treatment cleaning the SiO$_2$ surface. **(c)** An exfoliated graphene making conformal contact with the highly reactive SiO$_2$ surface, thus having high adhesion to the surface, this is achieved by directly exfoliating after the oxygen treatment. **(d)** An exfoliated graphene having less interaction with the SiO$_2$ surface due to the redeposition of environmental adsorbates from ambient exposure. **(e)** Areal percent of graphite flakes that survive the ultrasonic bath test, where each sample varies in the substrate's exposure time to ambient conditions. The dashed red line indicates a fitted line to the experimental data. Optical micrograph of an area of graphite **(f)** before and **(g)** after the ultrasound with high adhesion conditions. Optical micrograph of an area of graphite **(h)** before and **(i)** after the ultrasound with poor adhesion conditions.

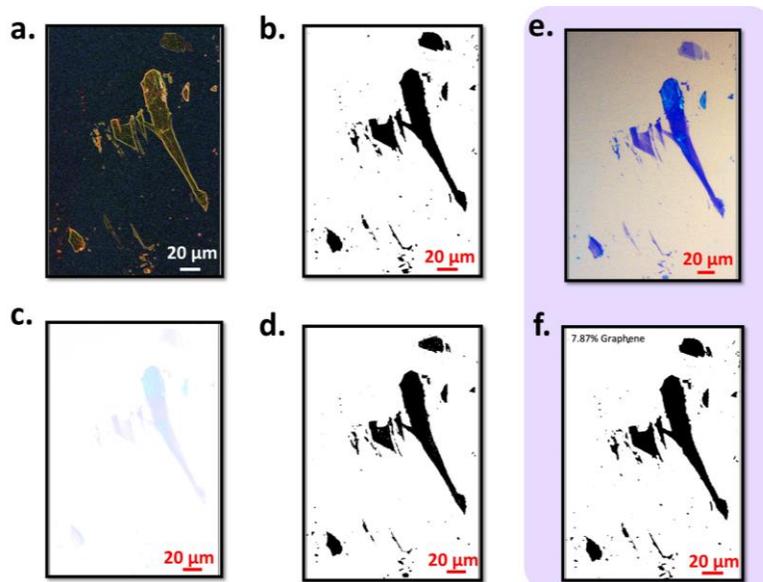

**Figure 2:** ImageJ analysis. (a,b) Method #1 of the image analysis, using an edge detection algorithm. **(a)** Shows the intensity gradient calculated from the original image (e), **(b)** shows the final binary image created by filling in the detected edges in (a). (c,d) Method #2 of the image analysis, using image thresholding. **(c)** The image's contrast is increased such that the background (substrate) is saturated, **(d)** shows the final binary image by applying a threshold to (c). **(e)** Original optical micrograph, **(f)** the final binary image which is an overlay of (b) and (d).

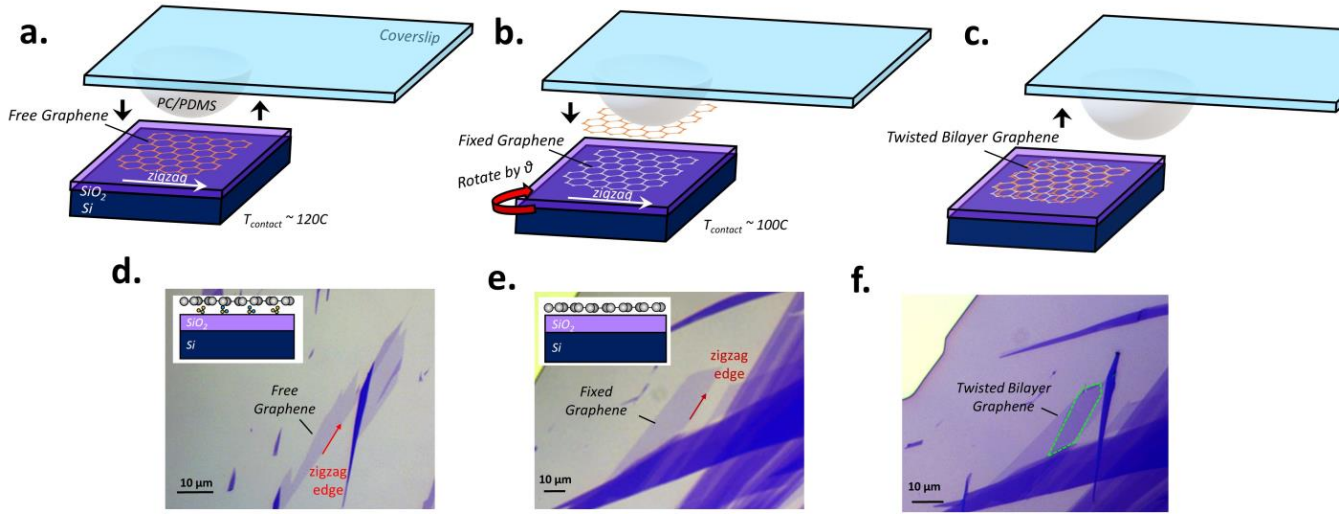

**Figure 3:** General TBG fabrication process. **(a)** PC/PDMS stamp hovering over a free graphene on 300 nm SiO$_2$/Si substrate. The stamp is brought down to pick up the free graphene layer. **(b)** The fixed graphene on a different substrate is placed underneath the stamp, with its zigzag edge matching the zigzag edge of the free graphene picked up from the previous step. The fixed graphene is rotated to the desired angle underneath the stamp, then the stamp is brought down to make contact between the two graphene monolayers. **(c)** The stamp is brought back up and will no longer have the "free" graphene attached, therefore leaving a TBG sample behind. **(d)** Optical micrograph a "free" graphene sample used with the zigzag edge identified. **(e)** Optical micrograph of the "fixed" graphene now underneath the stamp, with the zigzag edge matching the "free" graphene's edge. **(f)** Optical micrograph of the final fabricated TBG sample (dashed green outline), after removing the stamp.

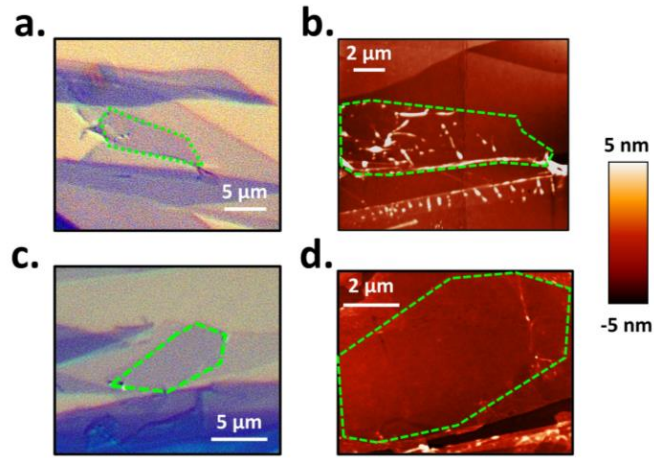

**Figure 4:** Optimal temperature when contacting the two graphene monolayers. **(a)** Differential interference contrast micrograph of a TBG sample when the contact temperature was conducted at 50°C, clearly displaying bubbles throughout the TBG region. **(b)** Atomic force micrograph of the same sample from (a), confirming the location of the bubbles. **(c)** Differential contrast micrograph of a TBG sample when the contact temperature was conducted at 100°C, where there is no clear observation of any bubbles in the sample. **(d)** Atomic force micrograph of the same sample from (c), confirming a clean and flat TBG region. Average surface roughness on TBG and $SiO_2$/Si individually match to be 0.09 nm.

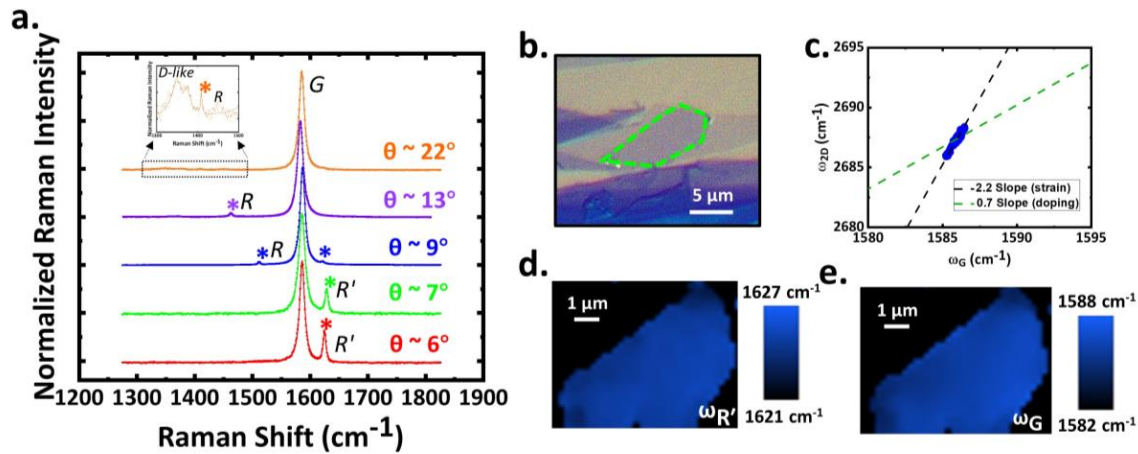

**Figure 5:** Raman spectroscopy on various fabricated TBG samples. **(a)** Raman spectra from successful TBG samples ranging in angles from 6-22°, confirmed via the R (R') - peak positions (asterisks are put above the R (R') peaks for reference). Inset is a zoomed in version of the D-like and R'-band arising from a larger angle sample of 22°. **(b)** Optical micrograph of a 6° TBG sample. **(c)** Plot of $\omega_{2D}$ versus $\omega_G$ from the TBG sample in (b). **(d)** Raman map of the R'-band position in the TBG sample from (b). **(e)** Raman map of the G-band position in the TBG sample from (b). Each Raman map has the corresponding scale bar exactly to the right of the map.

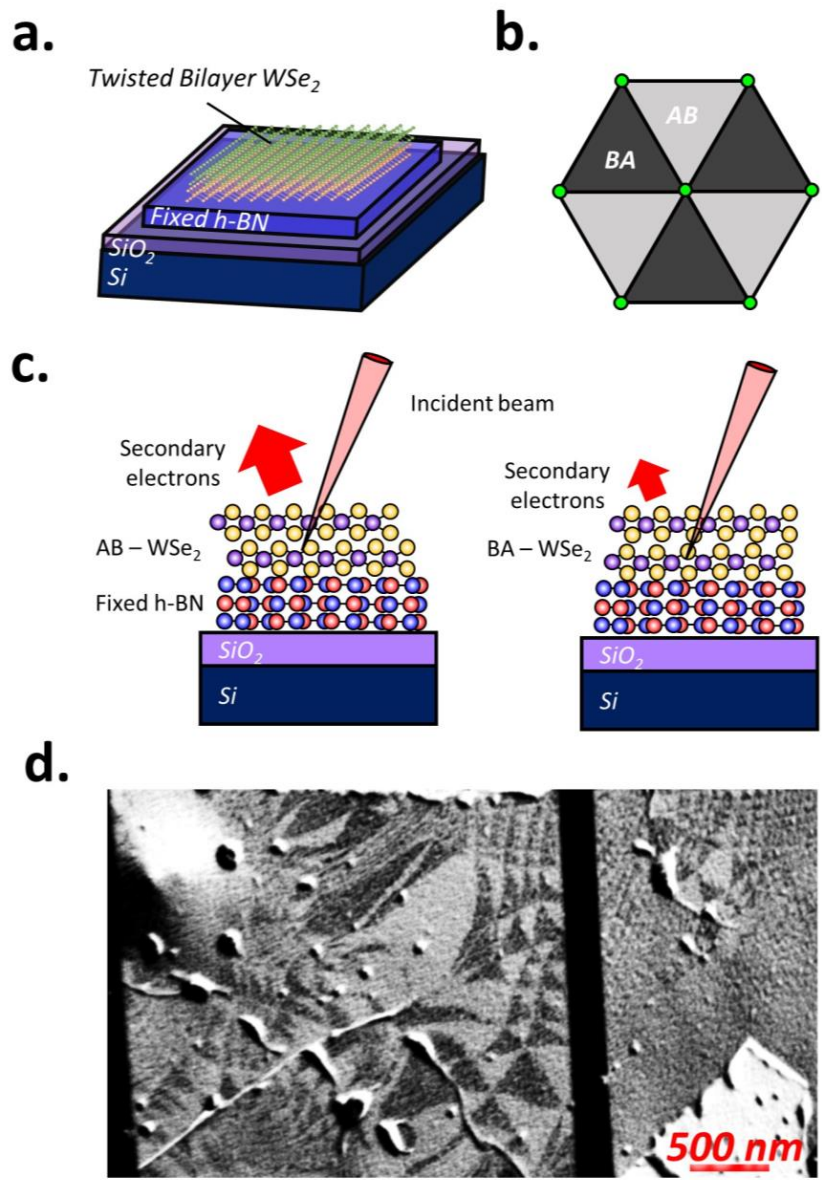

**Figure 6: (a)** Schematic of a marginally twisted bilayer WSe$_2$/h-BN/SiO$_2$/Si structure. **(b)** Stacking order diagram of a typical highly reconstructed bilayer moiré superlattice, where there are triangular AB and BA domains. Green points denote AA stacking regions, and the borders denote SP stacking. **(c)** Lattice channeling effect using scanning electron microscopy, where more secondary electrons are emitted when the incident beam is parallel to the specific stacking order. **(d)** Secondary electron image of a marginally twisted WSe$_2$/h-BN/SiO$_2$/Si structure utilizing this modified dry-transfer technique. White and black triangular domains indicate the reconstructed moiré AB/BA domains, showing a twist angle of close to 0°.

# Supplementary Material:

# Ultrasonic Delamination Based Adhesion Testing for High-Throughput Assembly of van der Waals Heterostructures

Tara Peña,*,† Jewel Holt,† Arfan Sewaket,† and Stephen M. Wu†,¶

†*Department of Electrical & Computer Engineering, University of Rochester, Rochester, NY, USA.*

¶*Department of Physics & Astronomy, University of Rochester, Rochester, NY, USA.*

**Ultrasonic Bath Test**

     For the ultrasonic bath test results presented in Fig. 1e of the main text, we image over the entire sample containing graphite flakes before and after the ultrasonic bath test (using a 50x objective lens). Only graphite flakes from monolayer to ~20 nm of thickness are imaged, since much thicker flakes tend to have more folds and wrinkles that are vulnerable to the ultrasonic bath. The $SiO_2$/Si substrates have markers engraved to ensure we may precisely reimage over the regions. Fig. 1e in the main text presents the mean value of the flakes that survive the ultrasonic bath, while the standard deviation is calculated from the variation between imaged regions of the respective sample. The ultrasonic bath's frequency is 40 kHz.

**Stamp Preparation**

     A 3 x 3 mm square Gel-Pak PDMS layer is placed on cleaned coverslips. To achieve a dome shaped PDMS layer, a droplet of Sylgard PDMS is placed onto the Gel-Pak PDMS square, then the entire coverslip is placed upside down to cure at room temperature for 24 hours. Chloroform with dissolved PC pellets is then spun on another cleaned glass slide. After 20 minutes, the chloroform will evaporate, leaving only the thin layer of PC on the cleaned glass slide. Then a double-sided Kapton tape window is created to gently grab the thin PC layer, then placed over the cured PDMS dome. Finally, the stamps are annealed at 120°C for 15 minutes, to have the PC adhere better to the PDMS dome and remove any potential nonuniformities in the film.

**Transfer Process**

     Flake-to-flake adhesion variation is quite common in exfoliated 2D materials; therefore, we ensure all of the high-adhered flakes can survive a 30-minute ultrasonic testing before we begin the transfer process. All transfers are then conducted inside an inert glovebox environment (<1 ppm $H_2O$ and $O_2$). Micro-manipulators are used to control for both the sample and stamp stages in X, Y, and Z directions. A ceramic heater is placed over a rotation stage, to allow for temperature and angle control over the substrates during the transfer. When the stamp is in contact with the desired monolayers, it is left in contact for minimally 5 minutes at the required temperature before cooling down for removal. A heating and cooling ramp rate is kept at 10°C per minute to attempt to control a contact speed of 1 μm/s throughout the entire transfer.

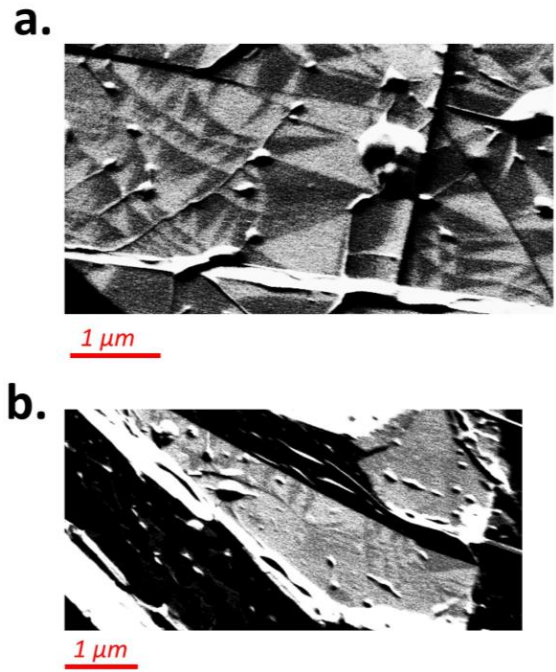

**Figure S1: (a,b)** Secondary electron images displaying of marginally twisted bilayer WSe$_2$/h-BN/SiO$_2$/Si samples (θ~0°). Roughly 500 nm AB/BA sized domains can be achieved with this method.